\documentclass[a4paper]{jpconf}
\usepackage{graphicx}
\usepackage[pass]{geometry}
\usepackage{graphicx}
\usepackage{amsmath}
\usepackage[utf8]{inputenc}
\usepackage{amsmath}
\usepackage{array}
\usepackage{amsfonts}
\usepackage{epsf}
\usepackage{epsfig}
\usepackage{amssymb}
\usepackage{bbm}
\usepackage{delarray}
\usepackage{subfig}

\begin{document}
\title{Feinberg-Horodecki Equation with Pöschl-Teller Potential: Space-like Coherent States}

\author{Altuğ Arda}

\address{Department of
Physics Education, Hacettepe University, 06800, Ankara, Turkey}

\ead{arda@hacettepe.edu.tr}

\author{Ramazan Sever}

\address{Department of
Physics, Middle East Technical  University, 06531, Ankara,Turkey}

\ead{sever@metu.edu.tr}

\begin{abstract}
We obtain the quantized momentum solutions, $\mathcal{P}_{n}$, of the Feinberg-Horodecki equation. We study the space-like coherent states for the space-like counterpart of the Schrödinger equation with trigonometric Pöschl-Teller potential which is constructed by temporal counterpart of the spatial Pöschl-Teller potential.
\end{abstract}

\section{Introduction}
It is well known that the basic equation for non-relativistic quantum mechanical systems is the Schrödinger equation which is a time-like equation writing for space-dependent potentials. This equation describes the dynamics of quantal systems. Actually, the time- and space-like equations are symmetric according to time and spatial coordinates, and it is possible to construct a "generalized quantum theory" including the above space-like quantum states [1]. Such a relativistic theory has been introduced by Feinberg [2], and it's non-relativistic version has been obtained by Horodecki [3] who wrote the following space-like counterpart of the Scrödinger equation in one-dimension
\begin{eqnarray}
\left(i\hbar\,\frac{\partial}{\partial x}+\frac{\hbar^2}{2m_{0}c}\,\frac{\partial^2}{\partial x_{0}^{2}}-\frac{V}{c}\right)\Psi=0\,,
\end{eqnarray}
which is the Feinberg-Horodecki equation. Here, $V$ denotes a vector potential, $m_{0}$ is the mass of particle and $x_{0}=ct$ with the speed of light $c$. By considering $\Psi=\psi(t)e^{-i\mathcal{P}_{n}x/\hbar}$, the space-independent form of this equation is written as [1]
\begin{eqnarray}
\left\{\frac{d^2}{dt^2}+\frac{c}{M(c)}\mathcal{P}-\frac{1}{M(c)}V(t)\right\}\psi(t)=0\,.
\end{eqnarray}
for stationary states with quantized momentum $\mathcal{P}_{n}$ ($n=0, 1, 2, \ldots$), where $M(c)=\hbar^2/2m_{0}c^2$.

The space-like quantum systems with the Feinberg-Horodecki equation receive a great attention especially in some branches of physics, such as in extended special relativity and in extended quantum mechanics [1, 4-6]. For example, they are used to explain the mass in a stable particle, the source of electric charge, and the force between electric charges [1, 7]. Among these works, Molski has also constructed the space-like coherent states of a time-dependent Morse potential with the help of Feinberg-Horodecki equation and showed that the obtained results for space-like coherent states can be used for Gompertzian systems [1]. Recently, only the quantized momentum solutions of the Feinberg-Horodecki equation for different time-dependent exponential potentials have been studied in literature [8, 9]. In the present work, we study the solutions of the Feinberg-Horodecki equation and extend the subject of coherent states to the space-like coherent states for the temporal counterpart of the Pöschl-Teller potential [10].

In view of practical ground, the P\"{o}schl-Teller potential is a more reliable approximation of quantum dots confinement potential, especially in axial direction [11]. The quantum wells having nonlinear optical properties (optical rectification, electrooptic effect and so on) provide useful opportunities for device applications in photodetectors, etc. It seems that P\"{o}schl-Teller potential has a successful use in semiconductors because of its tunable asymmetry degree [12]. Recent results based on nanophysics show that it is needed to construct the quantum states for infinite wells and these can be built by P\"{o}schl-Teller potential. Another property of it is that the P\"{o}schl-Teller potential has a rich quadratic spectrum structure for its coherent states [13]. Within the experimental solid state physics, spin dependent effects in a neutron beam with opposite spins have been explained by taking the magnetic field strength as a form of P\"{o}schl-Teller potential [14]. Within the theoretical applications, it is clear that many other potentials can be obtained from P\"{o}schl-Teller potential by some transformation and limiting procedures [13], so this potential is an interesting one in path-integral formalism serving as a testing ground for new evaluaiton methods [15].

The coherent states within quantum mechanics have been first constructed by Schrödinger in 1926 for the harmonic oscillator [16]. Later, Klauder-Glauber-Sudarshan have studied the coherent states as appropriate one for description of intense beam of photons [17]. Barut and Girardello have proposed another definition for coherent states (Barut-Girardello coherent states) which are the eigenstates of a lowering (annihilation) group operator with complex eigenvalues $z$ [18, 19]. The coherent states features that the expectation values of momentum and position are described by the classical equations of motion of a harmonic oscillator. The coherent states for the oscillator can be seen as the most classical of states which include also the squeezed states [20]. Nowadays, the coherent and squeezed states within the context of the continuous wave packets (as nonclassical states) become a fundamental point in quantum information theory and quantum computation. It seems also that the nonclassical states are a milestone for modern quantum optics [21, 22].

The study of coherent states for harmonic oscillator and other types of potentials has received great attention in literature, such as the coherent states for power-law potentials [23, 24], Morse potential [25, 26], pseudoharmonic oscillator [27], the coherent states with Kepler-Coulomb problem [20], and Pöschl-Teller potential and it's different forms [10, 28-31]. In this letter, we study the coherent states based on the definition given by Barut and Girardello of a space-like Pöschl-Teller potential which can be constructed by the temporal counterpart of the spatial Pöschl-Teller potential. In the present work, we tend to discuss the coherent states for a particular quantum system with the underlying symmetry. So, this study maybe bring more insights about the space-like quantum systems within the quantum mechanics.

The organization of this work is as follows. In the first part of Section 2, we present the quantized momentum values with the corresponding normalized solutions for the time-dependent Pöschl-Teller potential. We obtain briefly the lowering and raising operators including the Casimir operator, which belong to $su(1, 1)$ algebra, of the system under consideration. In next subsection, we study the space-like coherent states according to definition of Barut-Girardello. We give our conclusions in Section 3.

\section{Feinberg-Horodecki Equation with Pöschl-Teller Potential}

\subsection{Solutions}
Taking the space-like Pöschl-Teller potential as [10]
\begin{eqnarray}
V(t)=\frac{A(A-1)}{\cos^{2}(c_{1}t)}\,,
\end{eqnarray}
and using a dimensionless parameter $\tau=c_{1}t$, Eq. (2) gives us
\begin{eqnarray}
\left\{c^{2}_{1}\frac{d^2}{d\tau^{2}}+\frac{c}{M(c)}\mathcal{P}-\frac{1}{M(c)}\frac{a(a-1)}{\cos^{2}(\tau)}\right\}\psi(t)=0\,.
\end{eqnarray}
For finding out the quantized momentum $\mathcal{P}_{n}$, we define a new variable $s=\frac{1}{2}(1-\sin(\tau))$, and writing the wavefunction as $\psi(s)=(1-s)^{p}s^{q}\phi(s)$ in (4)
\begin{eqnarray}
s(1-s)\frac{d^2\phi(s)}{ds^2}+\left[2p+\frac{1}{2}-(1+4p)s\right]\frac{d\phi(s)}{ds}+\left(-4p^2+\frac{c\mathcal{P}_{n}}{c_{1}^{2}M(c)}\right)\phi(s)=0\,.
\end{eqnarray}
The last equation is then a hypergeometric-type equation if the parameters satisfy
\begin{eqnarray}
p=q=\frac{A'}{4}\,\,\,;\,\,\,\,\,\, A'=1+\sqrt{1+\frac{16A(A-1)}{c^{2}_{1}M(c)}\,}\,.
\end{eqnarray}
Comparing Eq. (5) with the hypergeometric equation [33]
\begin{eqnarray}
x(1-x)\frac{d^2\phi(x)}{dx^2}+[c-(a+b+1)x]\frac{d\phi(x)}{dx}-ab\phi(x)=0\,,
\end{eqnarray}
we write the solution of Eq. (5) in terms of the hypergeometric functions $\,_{2}F_{1}(a, b; c; x)$
\begin{eqnarray}
\phi(s) \sim \,_{2}F_{1}(a, b; c; s)\,,
\end{eqnarray}
with
\begin{eqnarray}
a=\frac{1}{2}\left[A'-\sqrt{\frac{4c\mathcal{P}_{n}}{c^{2}_{1}}\,}\,\right]; b=\frac{1}{2}\left[A'+\sqrt{\frac{4c\mathcal{P}_{n}}{c^{2}_{1}}\,}\,\right];c=\frac{1}{2}(1+A')\,.
\end{eqnarray}
In order to obtain a physical solution, we write $a=-n$ which gives the quantized momentum
\begin{eqnarray}
\mathcal{P}_{n}=\frac{c^{2}_{1}M(c)}{c}\left(n+\frac{A'}{2}\right)^{2}\,.
\end{eqnarray}
By using the above condition, we obtain the wavefunctions corresponding to quantized momentum of the system
\begin{eqnarray}
\psi(\tau)=N2^{-A'/2}\cos^{A'/2}(\tau)\,_{2}F_{1}(-n,n+A';\frac{1}{2}+\frac{A'}{2};\frac{1-\sin\tau}{2})\,.
\end{eqnarray}

For the normalization constant in Eq. (11), we change the variable as $y=1-2s$ ($0<y<1$), and use the relation between the hypergeometric functions and the associate Legendre polynomials $P_{k}^{\ell}(x)$ [33]
\begin{eqnarray}
P_{k}^{\ell}(x)=\frac{(-1)^{\ell}\Gamma(\ell+k+1)}{2^{\ell}\Gamma(k-\ell+1)\ell!}\,(1-x^2)^{\ell/2}\,_{2}F_{1}(\ell-k,\ell+k+1;\ell+1;\frac{1-x}{2})\,,
\end{eqnarray}
which gives us from Eq. (11)
\begin{eqnarray}
\psi(y)=N\,\frac{\Gamma(n+1)L!}{\sqrt{2}(-1)^{L}\Gamma(n+A')}\,(1-y^2)^{1/4}P_{n+L}^{L}(y)\,.
\end{eqnarray}
Writing the normalization condition as $\int_{0}^{\pi/2}|\psi(t)|^2dt=1$ gives the normalized wavefunctions
\begin{eqnarray}
\psi_{n}^{L}(y)=N_{n}^{L}(1-y^2)^{1/4}P_{K}^{L}(y)\,\,\,; N_{n}^{L}=\sqrt{\frac{(2n+2L+1)\Gamma(n+1)}{\Gamma(n+2L+1)}\,}\,.
\end{eqnarray}
where $L=-1/2+A'/2$, $K=n+L$, and used the orthogonality equation for the assosicate Legendre polynomials [32]
\begin{eqnarray}
\int^{1}_{0}P_{\ell}^{m}(x)P_{\ell'}^{m}(x)dx=\,\frac{1}{2\ell+1}\,\frac{(\ell+m)!}{(\ell-m)!}\,\delta_{\ell\ell'}\,.
\end{eqnarray}
and we take $c_1 \rightarrow 1$ for simplicity.

We are now ready to find the raising and lowering operators for our system which are needed to study the Barut-Girardello coherent states. For this aim, we use the following expression for associate Legendre polynomials [33]
\begin{eqnarray}
(1-x^2)\frac{dP^{\mu}_{\nu}}{dx}=(\nu+1)P^{\mu}_{\nu}-(\nu-\mu+1)P^{\mu}_{\nu+1}\,,
\end{eqnarray}
which gives
\begin{eqnarray}
\left[-(1-y^2)\frac{d}{dy}+y(n+L+\frac{1}{2})\right]\psi_{n}^{L}=\frac{N_{n}^{L}}{N_{(n+1)+L}^{L}}\,(n+1)\psi_{n+1}^{L}\,,
\end{eqnarray}
introducing the explicit form of the normalization constant
\begin{eqnarray}
\left[-(1-y^2)\frac{d}{dy}+y(n+L+\frac{1}{2})\right]\sqrt{\frac{2n+2L+1}{2n+2L+3}\,}\psi_{n}^{L}=\sqrt{(n+1)(n+2L+1)\,}\psi_{n+1}^{L}\,.
\end{eqnarray}
The left hand side of the equation can be identified as the raising operator based on the Feinberg-Horodecki equation for the time-dependent Pöschl-Teller potential
\begin{eqnarray}
\Gamma^{+}=\left[-(1-y^2)\frac{d}{dy}+y(n+L+\frac{1}{2})\right]\sqrt{\frac{2n+2L+1}{2n+2L+3}\,}\,.
\end{eqnarray}
Using the identity for the associate Legendre polynomials [33]
\begin{eqnarray}
(1-x^2)\frac{dP^{\mu}_{\nu}}{dx}=-\nu xP^{\mu}_{\nu}+(\nu+\mu)P^{\mu}_{\nu-1}\,,
\end{eqnarray}
we write
\begin{eqnarray}
\left[(1-y^2)\frac{d}{dy}+\frac{3y}{2}\right]\psi_{n}^{L}=\frac{N_{n}^{L}}{N_{(n-1)+L}^{L}}\,(n+2L)\psi_{n-1}^{L}\,,
\end{eqnarray}
and by writing the explicit form of the normalization constant we obtain
\begin{eqnarray}
\left[(1-y^2)\frac{d}{dy}+\frac{3y}{2}\right]\sqrt{\frac{2n+2L+1}{2n+2L-1}\,}\psi_{n}^{L}=\sqrt{n(n+2L)\,}\psi_{n-1}^{L}\,.
\end{eqnarray}
So, we give the lowering operator of our system
\begin{eqnarray}
\Gamma^{-}=\left[(1-y^2)\frac{d}{dy}+\frac{3y}{2}\right]\sqrt{\frac{2n+2L+1}{2n+2L-1}\,}\,.
\end{eqnarray}
We summarize the eigenvalue equations of the adjoint operators as
\begin{eqnarray}
\Gamma^{+}\psi_{n}^{L}&=&\sqrt{(n+1)(n+2L+1)\,}\psi_{n+1}^{L}\,,\nonumber\\
\Gamma^{-}\psi_{n}^{L}&=&\sqrt{n(n+2L)\,}\psi_{n-1}^{L}\,.
\end{eqnarray}
With the help of Eq. (24), it is easy to write the commutator
\begin{eqnarray}
\left[\Gamma^{-},\Gamma^{+}\right]\psi_{n}^{L}=2\Gamma_{0}\psi_{n}^{L}\,,
\end{eqnarray}
with the operator [10]
\begin{eqnarray}
\Gamma=\hat{n}+L+\frac{1}{2}\,,
\end{eqnarray}
where we used the eigenvalue as $\Gamma_{0}=n+L+1/2$ [34]. It is also possible to give the commutation relations between the above operators as
\begin{eqnarray}
\left[\Gamma,\Gamma^{+}\right]&=&+\Gamma^{+}\,,\nonumber\\
\left[\Gamma,\Gamma^{-}\right]&=&-\Gamma^{-}\,,
\end{eqnarray}
which means that the operator algebra of these operators belongs to $su(1, 1)$ algebra. Finally, the operator with the following action on the states $\psi_{n}^{L}$
\begin{eqnarray}
\left[\Gamma^2-\frac{1}{2}\left(\Gamma^{-}\Gamma^{+}+\Gamma^{+}\Gamma^{-}\right)\right]\psi_{n}^{L}=\tilde{L}(\tilde{L}-1)\psi_{n}^{L}\,,
\end{eqnarray}
can be identified as the Casimir operator
\begin{eqnarray}
C=\Gamma^2-\frac{1}{2}\left(\Gamma^{+}\Gamma^{-}+\Gamma^{-}\Gamma^{+}\right)\,.
\end{eqnarray}
with the eigenvalue $\tilde{L}=L-1/2$ [10].

In the next section, we study the Barut-Girardello coherent states for time-dependent Pöschl-Teller potential based on the solutions of Feinberg-Horodecki equation and some basic points related with coherent states such as normalization, it's orthogonality and expectation values of a physical observable $\mathcal{O}$ with respect to the coherent states.

\subsection{Barut-Girardello Coherent States}

According to Barut and Girardello [18], the coherent states are written as the eigenstates of the lowering operator $\Gamma^{-}$
\begin{eqnarray}
\Gamma^{-}|z,L>=z|z,L>\,,
\end{eqnarray}
where $z$ is a complex number. In the rest of computation, we denote the eigenstates of the temporal counterpart of the Pöschl-Teller potential as $\psi_{n}^{L} \equiv |n,L>$ which construct a complete orthonormal basis
\begin{eqnarray}
<n,L|n',L>=\delta_{nn'}\,\,\,;\,\,\,\,\,\sum_{n=0}^{\infty}|n,L><n,L|=1\,.
\end{eqnarray}

The eigenstates $|z,L>$ can be represended as the superposition of the above complete orthonormal set
\begin{eqnarray}
|z,L>=\sum_{n=0}^{\infty}<n,L|z,L>|n,L>\,.
\end{eqnarray}
Acting the lowering operator on $|z,L>$ with Eq. (30), using Eq. (24) and orthonormalization in Eq. (31) gives
\begin{eqnarray}
<n,L|z,L>=\frac{z}{\sqrt{n(n+2L)\,}}<n-1,L|z,L>\,,
\end{eqnarray}
which turns into
\begin{eqnarray}
<n,L|z,L>=z^{n}\,\sqrt{\frac{\Gamma(2L+1)}{n!\Gamma(n+2L+1)}\,}\,<0,L|z,L>\,,
\end{eqnarray}
where used a recurrence procedure.

From the normalization condition for coherent states, $<z,L|z,L>=1$, we obtain
\begin{eqnarray}
<0,L|z,L>=\,\sqrt{\frac{|z|^{2L}}{I_{2L}(2|z|)\Gamma(2L+1)}\,}\,,
\end{eqnarray}
where $I_{m}(x)$ is the modified Bessel function of order $m$ satisfying [27]
\begin{eqnarray}
\sum_{n=0}^{\infty}\,\frac{x^n}{n!\Gamma(n+m+1)}=\,\frac{1}{x^m}I_{m}(2x)\,.
\end{eqnarray}

The Barut-Girardello coherent states for a potential which is the temporal counterpart of the Pöschl-Teller potential with a normalization factor are written as
\begin{eqnarray}
|z,L>=\,\sqrt{\frac{|z|^{2L}}{I_{2L}(2|z|)}\,}\sum_{n=0}^{\infty}\,\frac{z^n}{n!\Gamma(n+2L+1)}\,|n,L>\,.
\end{eqnarray}

Let us now discuss briefly the appropriate measure $d\sigma(z,L)$ by which the resolution of the identity is realized for the coherent states $|z,L>$ [22, 24]
\begin{eqnarray}
\int d\sigma(z,L)|z,L><z,L|=1\,.
\end{eqnarray}
Here, the complex variable $z$ can be written in polar coordinates as $z=re^{i\theta}$ and the related integrals are performed over the whole complex plane which means $0\leq \theta \leq 2\pi$, $0\leq r < \infty$. It is possible to suggest that [25, 27]
\begin{eqnarray}
d\sigma(z,L)=\frac{2}{\pi}I_{2L}(2|z|)K_{2L}(2|z|)rdrd\theta\,,
\end{eqnarray}
where the functions $K_{2L}(2|z|)$ are the modified Bessel function of the second kind. By using the following equation [33]
\begin{eqnarray}
\int_{0}^{\infty}x^{\mu}K_{\nu}(ax)dx=\frac{2^{\mu-1}}{a^{\mu+1}}\Gamma\left(\frac{\mu+\nu+1}{2}\right)\Gamma\left(\frac{\mu-\nu+1}{2}\right)\,.
\end{eqnarray}
could be showed that Eq. (38) is satisfied. For the last identity, $\text{Re}\,(\mu+1\pm \nu)>0$ and $\text{Re}\,a>0$ should be noted.

In order to complete the discussion, we want to give the expectation value of a physical observable $\mathcal{O}$ with respect to the Barut-Girardello coherent states which are based on the solutions of the Feinberg-Horodecki equation. With the help of (37), the expectation value of a physical observable $\mathcal{O}$ can be given as
\begin{eqnarray}
<z,L|\mathcal{O}|z,L>&\equiv& <\mathcal{O}>\nonumber\\&=&\frac{|z|^{2L}}{I_{2L}(2|z|)}\sum_{n,n'=0}^{\infty}\,\frac{(z^{*})^{n'}z^{n}}{\sqrt{n'!n!\Gamma(n'+2L+1)\Gamma(n+2L+1)\,}}
<n,L|\mathcal{O}|n,L>\,,\nonumber\\
\end{eqnarray}
for which one needs to evaluate a sum as following [24]
\begin{eqnarray}
\sum_{\ell=0}^{\infty}\,\frac{(x^2)^{\ell}}{\ell!\Gamma(\ell+m+1)}\ell^{n}\,.
\end{eqnarray}
For example, we have with $n=0$
\begin{eqnarray}
\sum_{\ell=0}^{\infty}\,\frac{x^{\ell}}{\ell!\Gamma(\ell+m+1)}=\frac{1}{x^{m}}I_{m}(2x)\,.
\end{eqnarray}

With this result, we complete our discussion about the space-like coherent states which based on the definition of Barut and Girardello for the temporal counterpart of the Pöschl-Teller potential with the solutions of the Feinberg-Horodecki equation.

\section{Conclusion}
We have extended the subject of coherent states to the space-like coherent states which are based on the solutions of the space-like counterpart of the Schrödinger equation called as Feinberg-Horodecki equation, namely. We have constructed the space-like coherent states according to definition of Barut-Girardello for a time-dependent Pöschl-Teller potential and discussed some basic points related with the coherent states such as normalization, the resolution of the identity satisfied by the coherent states, and possibility to find the expectation value of a physical observable $\mathcal{O}$.

\section{Acknowledgments}
This research was supported by Hacettepe University Scientific Research Coordination Unit, Project Code: FBB-$2016$-$9394$.

\end{document}